\documentclass[11pt]{article}

\usepackage{fullpage,latexsym,graphicx,url,amsmath,epsfig,epsf,latexsym}

\usepackage{cite}

\begin{document}

\title{ \vspace{-3cm}
 ~{\normalsize\rightline{preprint LA-UR-07-0471}}\\~\\~\\
Optimal Shape of a Blob}

\author{Carl M. Bender\thanks{Permanent address: Department of Physics,
Washington University, St. Louis MO 63130, USA.}\\
\footnotesize Center for Nonlinear Studies\\
\footnotesize Los Alamos National Laboratory\\
\footnotesize Los Alamos, NM 87545, USA
\\\footnotesize\texttt{cmb@wustl.edu}
\and ~
\and Michael A. Bender\\
\footnotesize Department of Computer Science\\
\footnotesize Stony Brook University\\
\footnotesize Stony Brook, NY 11794-4400, USA\\
\footnotesize\texttt{bender@cs.sunysb.edu} }

\date{\today}
\maketitle

\newcommand{\secput}[2]{\section{#2}\label{sec:#1}}
\newcommand{\subsecput}[2]{\subsection{#2}\label{sec:#1}}

\newcommand{\figlabel}[1]{\label{fig:#1}}

\newcommand{\secref}[1] {Sec.~\protect\ref{sec:#1}}
\newcommand{\secreftwo}[2] {Sections~\protect\ref{sec:#1} and~\protect\ref{sec:#2}}
\newcommand{\secrefthree}[3] {Sections \ref{sec:#1}, \ref{sec:#2}, and \ref{sec:#3}}
\newcommand{\secreffour}[4] {Sections \ref{sec:#1}, \ref{sec:#2}, \ref{sec:#3}, and~\ref{sec:#4}}
\newcommand{\appref}[1] {Appendix~\ref{app:#1}}
\newcommand{\figref}[1] {Fig.~\protect\ref{fig:#1}}
\newcommand{\figreftwo}[2] {Figs.~\protect\ref{fig:#1} and~\protect\ref{fig:#2}}
\newcommand{\tabref}[1] {Table~\ref{tab:#1}}
\newcommand{\eqlabel}[1] {\label{eq:#1}}
\renewcommand{\eqref}[1] {(\protect\ref{eq:#1})}
\newcommand{\eqreftwo}[2] {(\protect\ref{eq:#1})~and~(\protect\ref{eq:#2})}


\let\latexcite=\cite
\def\cite{\nolinebreak\latexcite}
\let\latexref=\ref
\def\ref{\nolinebreak\latexref}

\newcommand{\ginv}{g^{-1}}
\newcommand{\hinv}{h^{-1}}

\begin{abstract}
This paper presents the solution to the following optimization problem: What is
the shape of the two-dimensional region that minimizes the average $L_p$
distance between all pairs of points if the area of this region is held fixed?
[The $L_p$ distance between two points ${\bf x}=(x_1,x_2)$ and ${\bf y}=(y_1,y_2
)$ in $\Re^2$ is $\left(|x_1-y_1|^p+|x_2-y_2|^p\right )^{1/p}$.] Variational
techniques are used to show that the boundary curve of the optimal region
satisfies a nonlinear integral equation. The special case $p=2$ is elementary
and for this case the integral equation reduces to a differential equation whose
solution is a circle. Two nontrivial special cases, $p=1$ and $p=\infty$, have
already been examined in the literature. For these two cases the integral
equation reduces to nonlinear second-order differential equations, one of which
contains a quadratic nonlinearity and the other a cubic nonlinearity.
\end{abstract}

\secput{intro}{Introduction}

A general class of optimization problems is stated as follows: Given a set of
$n$ points in a metric space and an integer $k<n$, find a subset of size $k$
that minimizes or maximizes the average distance between all pairs of points in
the subset. Because this problem is NP-complete~\cite{Krumke97}, it is believed
that there is no algorithm that solves this problem in time polynomial in $n$
(or $k$).

This class of optimization problems has been widely discussed in the computer
science literature. One possible application for which the average distance is
minimized is in the allocation of jobs to nodes in a
supercomputer~\cite{Krumke97,MacheLoWi97,MacheLo97,Leung02,BenderBDFLMP05}.
Another possible application is in VLSI layout, where the objective is to
cluster nearby components of a circuit on a chip~\cite{AhmadiniaBFTV}. This
class of problems is general enough to model physical problems, where one seeks
the lowest-energy state of a set of particles having pairwise attractive or
repulsive forces.

Even special cases have subtle computational-complexity issues. For example,
suppose that one is given a set of $n$ \emph{integer grid points} and the
objective is to find a set of $k$ points that minimizes the average pairwise
Manhattan or Euclidean distance between points. It is not known if these special
cases are NP-complete, and no polynomial-time algorithms have been discovered.
Furthermore, even if the original $n$ points make up a square region of the grid
it is not known if this problem is NP-complete. This family of problems is so
computationally difficult that the usual approach in the literature has been to
find algorithms that produce approximate solutions~\cite{Hochbaum97}.

This paper presents the solution to a continuum version of an optimization
problem from this class of discrete problems. To be specific, our limited
objective here is to identify and to characterize the optimal two-dimensional
geometrical shape of a blob that minimizes the average pairwise distance between
points in the blob. An intuitive way to understand the optimal shape is to view
think of the blob as a city. By optimal, we mean that the average travel
distance between any two points in the city is minimized. For the case of a
city, a particularly appropriate metric is the \emph{Manhattan distance}, which
is the sum of the east-west distance along streets and the north-south distance
along avenues. It is interesting that while the optimal shape of a city is not
circular, it is extremely close to circular~\cite{KarpMcWo75,BenderBeDeFe04}.

In this paper we consider the general case of $L_p$ norms in $\Re^2$, for which
the $L_p$ distance between two points ${\bf x}=(x_1,x_2)$ and ${\bf y}=(y_1,
y_2)$ is defined to be
\begin{equation}
||{\bf x}-{\bf y}||_p\equiv\big(|x_1-y_1|^p+|x_2-y_2|^p\big)^{1/p}\quad(p>1).
\eqlabel{e1}
\end{equation}
We use variational methods to determine the shape of the region that minimizes
the average $L_p$ distance when the area of this region is held fixed.
Specifically, we characterize the boundary curve $h(x)$ of the optimal region
and show that this boundary curve satisfies a nonlinear integral equation. This
integral equation is new and is the principal result of this paper.

This paper generalizes two earlier studies. Karp, McKellar, and
Wong~\cite{KarpMcWo75} consider the two special extreme cases $p=1$ and $p=
\infty$. For these cases, they determine a differential equation that describes
the boundary of the optimal region that minimizes the average distance between
all points in the region. Bender et al.~\cite{BenderBeDeFe04} study the
dimensionless average pairwise distance $D[h]$ that characterizes the degree of
optimality of the region for $p=1$ and they use variational methods to minimize
the value of this functional $D[h]$. This variational approach leads directly to
the nonlinear differential equation that describes the boundary curve. In
Ref.~\cite{BenderBeDeFe04} the numerical value of $D[h]$ is also computed.

Here we extend the variational methods introduced in Ref.~\cite{BenderBeDeFe04}
to the general case $p\geq1$. For this situation the boundary curve satisfies
an integral equation. We then examine three special cases: For $p=2$ (the
Euclidean metric) the integral equation reduces to a separable first-order
nonlinear differential equation, whose solution, as one would expect, is a
circle. For $p=1$ (the Manhattan metric) and for $p=\infty$ (the \emph{maximum}
or \emph{Chebychev} metric) the integral equation reduces to the second-order
nonlinear differential equations that are previously derived in
Refs.~\cite{KarpMcWo75,BenderBeDeFe04}.

This paper is organized as follows: In \secref{formulation} we show how to
derive the integral equation that defines the boundary of the optimal region for
general $p\geq1$. Then in \secref{special-values} we examine this integral
equation for the special cases $p=1$, $p=2$, and $p=\infty$.

\secput{formulation}{Derivation of the Integral Equation}

In this section we first formulate the optimization problem and then perform a
variational calculation~\cite{MarsdenRa99,Mercier63} to obtain an integral
equation whose solution is the boundary of the optimal region.

\subsection*{Formulation of the continuum problem}

Let $w(x)$ be the upper boundary of the optimal region. Without loss of
generality we assume that $w(x)$ is positive. The $L_p$ distance defined in
\eqref{e1} is up-down and left-right symmetric in $\Re^2$.
Thus, the lower bound of the region is $-w(x)$ and $w(x)=w(-x)$. The boundary of
the optimal region is continuous, so there is a point $x=a>0$ at which the curve
$w(x)$ crosses the $x$-axis: $w(a)=w(-a)=0$.

Furthermore, the optimal region is symmetric about the $45^\circ$ lines $y=x$
and $y=-x$. This symmetry allow us to decompose the function $w(x)$ into two
functions, $w(x)=g(x)$ below the line $y=x$ and $w(x)=h(x)$ above the line
$y=x$:
\begin{eqnarray}
w(x)=\left\{\begin{array}{ll} h(x)&\quad(0\leq x\leq b),\\
g(x)&\quad(b\leq x \leq a), \end{array} \right.
\eqlabel{e2}
\end{eqnarray}
where $b$ marks the point on the $x$-axis where the boundary curve $w(x)$
crosses the line $y=x$. The symmetry about the line $y=x$ implies that $h(x)$ is
the \emph{functional} inverse of $g(x)$: $h(x)=\ginv(x)$.

Finally, we define the length scale of this problem by choosing, without loss of
generality, to work in units such that $b=1$. We summarize the properties of the
functions $w(x)$, $h(x)$, and $g(x)$ as follows:
\begin{eqnarray}
w(0)&=&h(0)=a,\nonumber\\
w(1)&=&h(1)=g(1)=1,\nonumber\\
h(x)&=&g^{-1}(x)\quad(0\leq x\leq 1),\nonumber\\
w(x)&\geq&0\quad(-a\leq x\leq a).
\eqlabel{e3}
\end{eqnarray}
The functions $g(x)$ and $h(x)$ are illustrated in \figref{f00}.

\begin{figure}[eh!tb]
\centerline{~\includegraphics[height=2.5in]{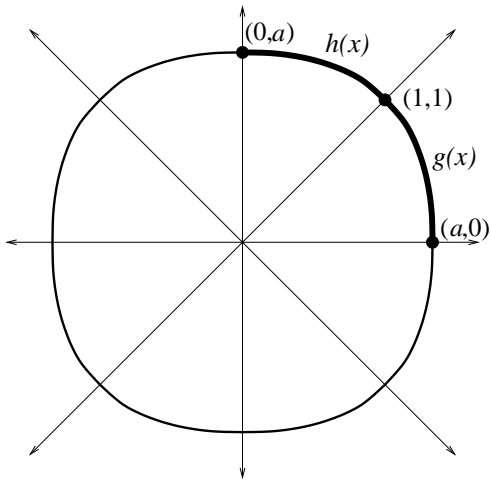}}
\caption{Notation used in this paper. The boundary of the optimal region is
symmetric with respect to reflections about the $x$ axis, the $y$ axis, and the
$45^\circ$ lines $y=\pm x$. In the upper-half plane the boundary curve is $w(x)$
[see \eqref{e3}]. The function $w(x)$ crosses the $y$ and $x$ axes at the points
$(0,a)$ and $(\pm a,0)$. The slope of $w(x)$ is $0$ at $x=0$ and infinite at $x=
a$. Also, $w(x)$ crosses the line $y=x$ at the point $(1,1)$, and at this point
its slope is $-1$. The portion of $w(x)$ in the range $0\leq x\leq1$ is $h(x)$
and the portion of $w(x)$ in the range $1\leq x\leq a$ is $g(x)$. The functions
$h(x)$ and $g(x)$ are inverses of one another because $w(x)$ is symmetric about
the line $y=x$.}
\figlabel{f00}
\end{figure}

The area of the optimal region is given by the functional $A[w]$ whose boundary
curve is shown in \figref{f00}. Note that $A[w]$ is four times the area in the
positive quadrant and eight times the area in the positive octant:
\begin{eqnarray}
A[w]&=&4\int_{x=0}^a\!\!\! dx\,w(x)\,,\nonumber\\
A[h]&=&8\int_{x=0}^1\!\!\! dx\,[h(x)-x]\,.
\eqlabel{e4}
\end{eqnarray}

We use the notation $M[w]$ to represent the integrated sum of the
distances between all pairs of points in the region:
\begin{equation}
M[w]=\int_{x=-a}^a\!\!\!dx\int_{y=-w(x)}^{w(x)}\!\!\!dy\int_{u=-a}^a\!\!\!du
\int_{v=-w(u)}^{w(u)}\!\!\!dv\,\left(|x-u|^p+|y-v|^p\right)^{1/p}\,.
\eqlabel{e5}
\end{equation}
By the symmetry of the boundary function $w(x)$, we may restrict the above
integration to the points $(x,y)$ in one of the octants:
\begin{equation}
\eqlabel{e6}
M[w]=8\int_{x=0}^1\!\!\! dx\int_{y=x}^{h(x)}\!\!\!dy\int_{u=-a}^a\!\!\!du
\int_{v=-w(u)}^{ w(u)}\!\!\!dv\,\left(|x-u|^p+|y-v|^p\right)^{1/p}\,.
\end{equation}
We adjust the integrands so that we integrate over positive regions only:
\begin{eqnarray}
M[w]&=&8\int_{x=0}^1\!\!\!dx\int_{y=x}^{h(x)}\!\!\!dy\int_{u=0}^a\!\!\!du
\int_{v=0}^{w(u)}\!\!\!dv\,\Big\{\left[|x-u|^p+|y-v|^p\right]^{1/p}+\left[|x-
u|^p+(y+v)^p\right]^{1/p}\nonumber\\
&&\qquad\qquad+\left[(x+u)^p+|y-v|^p\right]^{1/p}+\left[(x+u)^p+(y+v)^p\right]^{
1/p}\Big\}\,.
\eqlabel{e7}
\end{eqnarray}

Next, we use \eqref{e3} to replace the function $w$ with $h(u)$ and $g(u)$ to
obtain:
\begin{eqnarray}
M[w]=8\int_{x=0}^1\!\!\! dx\int_{y=x}^{h(x)}\!\!\!dy\int_{u=0}^1\!\!\!du
\int_{v=0}^{h(u)}\!\!\!dv\,\Big\{ \left[|x-u|^p+|y-v|^p\right]^{1/p}+\left[|x-
u|^p+(y+v)^p\right]^{1/p}\nonumber\\
+\left[(x+u)^p+|y-v|^p\right]^{1/p}+\left[(x+u)^p+(y+v)^p\right]^{1/p}\Big\}
\nonumber\\
+8\int_{x=0}^1\!\!\! dx\int_{y=x}^{h(x)}dy\int_{u=1}^a\!\!\!du\int_{v=0}^{g(u)}
\!\!\!dv\,\Big\{ \left[|x-u|^p+|y-v|^p\right]^{1/p}+\left[|x-u|^p+(y+v)^p
\right]^{1/p}\nonumber\\
+\left[(x+u)^p+|y-v|^p\right]^{1/p}+\left[(x+u)^p+(y+v)^p\right]^{1/p}\Big\}\,.
\eqlabel{e8}
\end{eqnarray}
In addition, we make a change of variables in \eqref{e8} to eliminate all
reference to the function $g$:\footnote{In this change of variables, $u=h(s)$,
$g(u)=s$, $u:1\rightarrow a$ becomes $s:1\rightarrow 0$.}
\begin{eqnarray}
M[h]=8\int_{x=0}^1 \!\!\!dx\int_{y=x}^{h(x)}\!\!\!dy\int_{u=0}^1\!\!\!du\int_{v=
0}^{h(u)}\!\!\!dv\, \Big\{ \left[(x+u)^p+|y-v|^p\right]^{1/p}+\left[|x-u|^p+|y-
v|^p\right]^{1/p}\nonumber\\
+\left[(x+u)^p+(y+v)^p\right]^{1/p}+\left[|x-u|^p+(y+v)^p\right]^{1/p} \Big\}
\nonumber\\
-8\int_{x=0}^1\!\!\! dx\int_{y=x}^{h(x)}\!\!\!dy\int_{s=0}^1\!\!\!ds\,h'(s)
\int_{v=0}^{s}\!\!\!dv\, \Big\{\left[(x+h(s))^p+|y-v|^p\right]^{1/p}+\left[(h(s)
-x)^p+|y-v|^p\right]^{1/p}\nonumber\\
+\left[(x+h(s))^p+(y+v)^p\right]^{1/p}+\left[(h(s)-x)^p+(y+v)^p\right]^{1/p}
\Big\}\,.
\eqlabel{e9}
\end{eqnarray}

\subsection*{Variational calculation}

We must find the function $h(x)$ that minimizes the functional $M[h]$ subject to
the constraint that the area $A[h]$ is held fixed. Since, the functionals $M[h]$
and $A[h]$ have units of $[\mbox{\sc length}]^5$ and $[\mbox{\sc length}]^2$,
respectively, a \emph{dimensionless} measure of the average pairwise distance is
given by the ratio $D[h]$:
\begin{equation}
D[h]=\frac{M[h]}{(A[h])^{5/2}}\,.
\eqlabel{e10}
\end{equation}
The functional $D[h]$ is a dimensionless measure of the optimality of the
region. We will now use variational methods to find the boundary curve $h$ that
minimizes the value of $D[h]$.

We begin by calculating the functional derivative of $D[h]$ with respect to
$h(t)$:
\begin{equation}
\frac{\delta D[h]}{\delta h(t)}=\frac{1}{(A[h])^{5/2}}\frac{\delta M[h]}{\delta
h(t)}-\frac{5}{2}\frac{M[h]}{(A[h])^{7/2}}\frac{\delta A[h]}{\delta h(t)}\,.
\eqlabel{e11}
\end{equation}
To evaluate the right side of this equation we calculate the functional
derivative of $M[h]$ in \eqref{e9}:
\begin{eqnarray}
\frac{\delta M[h]}{\delta h(t)}&=&16\int_{x=0}^{1}\!\!\!dx\left\{\int_{y=x}^{h
(x)}\!\!\!dy+\int_{y=0}^{x}\!\!\!dy\right\} \Big\{\left[(x+t)^p+|y-h(t)|^p
\right]^{1/p}+\left[|x-t|^p+|y-h(t)|^p\right]^{1/p}\nonumber\\
&&+\left[(x+t)^p+(y+h(t))^p\right]^{1/p}+\left[|x-t|^p+(y+h(t))^p\right]^{1/p}
\Big\}\nonumber\\
&&-16\int_{x=0}^{1}\!\!\!dx \, h'(x)\int_{y=0}^{x}\!\!\!dy\Big\{\left[(t+h(x))^p
+|h(t)-y|^p\right]^{1/p}+\left[(h(x)-t)^p+|h(t)-y|^p\right]^{1/p}\nonumber\\
&&+\left[(t+h(x))^p+(h(t)+y)^p\right]^{1/p}+\left[(h(x)-t)^p+(h(t)+y)^p\right]^{
1/p}\Big\}\,.
\eqlabel{e12}
\end{eqnarray}
We also calculate the functional derivative of the area $A[h]$ in \eqref{e4}:
$$\frac{\delta A[h]}{\delta h(t)}=8\,.$$

The above calculation expresses the functional derivative of $D[h]$ as a double
integral. However, after lengthy simplifications, we can reduce the double
integral to a single integral. Setting the result to 0 gives the integral
equation satisfied by the function $h$ that minimizes the functional $D[h]$:
\begin{eqnarray}
&&\quad\int_{x=0}^{1}\!\!\!dx\,\Big\{ \left[h'(x)+h'(t)\right]
\Big(|x-t|^{p}+\left[h(x)+h(t)\right]^{p}\Big)^{1/p}\nonumber\\
&&\!\!\!\!\!\!+\left[h'(x)-h'(t)\right]\Big(|x-t|^p+\left|h(x)-h(t)\right|^p
\Big)^{1/p}-\left[h'(x)+h'(t)\right]\Big((x+t)^p+\left|h(x)-h(t)\right|^p
\Big)^{1/p}\nonumber\\
&&\!\!\!\!\!\!-\left[h'(x)-h'(t)\right] \Big((x+t)^{p}+\left[h(x)+h(t)\right]^p
\Big)^{1/p}+\left[1+h'(x)h'(t)\right]\Big(\left[h(t)-x\right]^p
+\left[h(x)+t\right]^{p}\Big)^{1/p}\nonumber\\
&&\!\!\!\!\!\! +\left[1-h'(x)h'(t)\right]\Big(\left[h(t)+x\right]^p+\left[h(x)+t
\right]^p\Big)^{1/p}+\left[h'(x)h'(t)-1\right]\Big(\left[h(t)-x\right]^p+\left[h
(x)-t\right]^p\Big)^{1/p}\nonumber\\
&&\!\!\!\!\!\!-\left[1+h'(x)h'(t)\right]\Big(\left[h(t)+x\right]^p+\left[h(x)
-t\right]^p\Big)^{1/p} \Big\}=0\,.
\eqlabel{e13}
\end{eqnarray}
This equation is complicated, but we can rewrite it in a compact form by
changing the integration variables and the limits of integration. The result is
a nonlinear integral equation satisfied by the original function $w$:
\begin{eqnarray}
\int_{x=-a}^a\!\!\!dx\big[w'(x)+w'(t)\big]\Big[\big(|x-t|^p+\big[w(x)+w(t)
\big]^p\big)^{1/p}-\big(|x+t|^p+|w(x)-w(t)|^p\big)^{1/p}\Big]=0\,.
\eqlabel{e14}
\end{eqnarray}
The equivalent integral equations in \eqreftwo{e13}{e14}
are the principal result of this paper.

\secput{special-values}{Special Values of \boldmath$p$}

The integral equation \eqref{e14} reduces to differential equations
for three special values of $p$. Specifically, when $p=2$ it becomes
a first-order nonlinear differential equation, and for the extreme cases when
$p=1$ and $p=\infty$ it reduces to the second-order nonlinear differential 
equations derived in Refs.~\cite{KarpMcWo75,BenderBeDeFe04}.
In this section we investigate these three special cases.

\subsection*{Special case: \boldmath$p=2$}

If we set $p=2$ in \eqref{e14}, we can dispense with the absolute-value signs
and the integral equation now reads
\begin{eqnarray}
\int_{x=-a}^a\!\!\!dx\,\left[w'(x)+w'(t)\right]\left[\sqrt{(x-t)^2+\left[w
(x)+w(t)\right]^2}-\sqrt{(x+t)^2+\left[w(x)-w(t)\right]^2}\right]=0\,.
\eqlabel{e15}
\end{eqnarray}
It is not immediately obvious, but for this special case we can show that the
integrand in this integral equation becomes a total derivative if
$w(x)$ satisfies the differential equation
\begin{equation}
w'(x)w(x)+x=0\,.
\eqlabel{e16}
\end{equation}
Indeed, it is easy to verify that if \eqref{e16} holds, we can rewrite
\eqref{e15} in the form
\begin{eqnarray}
\int_{x=-a}^a\!\!\!dx\,\frac{\partial}{\partial x}\frac{1}{3w(t)}
\left[\Big((x-t)^2+\left[w(x)+w(t)\right]^2\Big)^{3/2}+
\Big((x+t)^2+\left[w(x)-w(t)\right]^2\Big)^{3/2}\right]=0\,.
\eqlabel{e17}
\end{eqnarray}

Evaluating this integral at the endpoints $x=\pm a$ and using the condition
that $w(\pm a)=0$ gives the value 0 for this integral. This verifies the
integral equation \eqref{e15}. Finally, solving \eqref{e16} for $w(x)$ subject
to the condition that $w(\pm1)=1$ gives the particular solution
\begin{eqnarray}
w(t)=\sqrt{2-t^2}.
\eqlabel{e18}
\end{eqnarray}
Thus, the boundary of the optimal region is a circle of radius $\sqrt{2}$.

Let us calculate the value of $D[h]$ for this circular region. For simplicity,
we calculate $M[h]$ for a circle of radius $1$ so that $h(x)=\sqrt{1-x^2}$ and
then divide by $(A[h])^{5/2}=\pi^{5/2}$. We calculate $D[h]$:
$$D[h]=\pi^{-5/2}\int_{x=-1}^1\!\! dx\,\int_{y=-\sqrt{1-x^2}}^{\sqrt{1-x^2}}dy\,
\int_{u=-1}^1\!\! du\,\int_{v=-\sqrt{1-u^2}}^{\sqrt{1-u^2}} dv\,\sqrt{(x-u)^2+
(y-v)^2}.$$
It is best to evaluate this integral in polar coordinates:
$$D[h]=\pi^{-5/2}\int_{r=0}^1 dr\,r\int_{\theta=0}^{2\pi} d\theta\,\int_{s=0}^1
ds\,s\int_{\phi=0}^{2\pi}d\phi\sqrt{(r\cos\theta-s\cos\phi)^2+(r\sin\theta-s\sin
\phi)^2}.$$

Next, we make the change of variable $s=rt$ and use circular symmetry to
evaluate the $\phi$ integral. We obtain
$$D[h]=2\pi^{-3/2}\int_{r=0}^1 dr\,r^4\int_{t=0}^{1/r} dt\,t\int_{\theta=0}^{2
\pi}d\theta\,\sqrt{1+t^2-2t\cos\theta}.$$
We interchange orders of integration and perform the $r$ integral exactly
to obtain
$$D[h]=\frac{4}{5}\pi^{-3/2}\int_{t=0}^1 dt\,t \int_{\theta=0}^{2\pi} d\theta\,
\sqrt{1+t^2-2t\cos\theta}.$$
The integrals with respect to $t$ and $\theta$ are easy to perform and the
exact result is
$$D[h]=\frac{16}{15}\pi^{-3/2}\approx0.1915596\,.$$

\subsection*{Special case: \boldmath $p=1$}

We now show that when $p=1$, the integral equation \eqref{e13} reduces to the
differential equation that was derived by two different methods in
Refs.~\cite{KarpMcWo75} and~\cite{BenderBeDeFe04}. 
We begin by setting $p=1$ in \eqref{e13} and simplify the result to obtain
\begin{eqnarray}
\int_{x=0}^{1}\!\!\!dx\Big(h'(x)\big(|x-t|-x-t\big)
+h'(t)\big[h(x)+h(t)-|h(x)-h(t)|\big]-2xh'(x)h'(t)+2t\Big)=0\,.
\eqlabel{e19}
\end{eqnarray}
Then, we perform several integrations by parts to obtain the simpler integral
equation
\begin{eqnarray}
\int_{x=0}^{t}\!\!\!dx\,h(x)-h'(t)+2h'(t)\int_{x=0}^{1}\!\!\!dx\,h(x)
-h'(t)\int_{x=0}^{t}\!\!\!dx\,h(x)+h'(t)h(t)t=0\,.
\eqlabel{e20}
\end{eqnarray}

We reduce this integral equation to a differential equation by
defining the new function $f(t)$, whose derivative is $h(t)$:
\begin{eqnarray}
f(t)\equiv \int_{u=0}^t\!\!\! dx\,h(x)\,.
\eqlabel{e21}
\end{eqnarray}
In terms of $f(t)$ the integral equation \eqref{e20} becomes
\begin{equation}
tf'(t)f''(t)+[2f(1)-1]f''(t)+f(t)-f''(t)f(t)=0\,.
\eqlabel{e22}
\end{equation}

We must specify the boundary conditions satisfied by $f(t)$.
By the definition in~\eqref{e21} we have
\begin{equation}
f(0)=0\,.
\eqlabel{e23}
\end{equation}
Also from \eqref{e3} we have
\begin{eqnarray}
f'(1)=h(1)=1\,.
\eqlabel{e24}
\end{eqnarray}
The solution to the differential equation \eqref{e22} is uniquely
determined by these boundary conditions. While an analytical
solution is not known, the numerical solution is given in
Ref.~\cite{BenderBeDeFe04}. Using this numerical solution,
it is shown that the numerical value of $D[h]$ is $0.650\,245\,952\,951$.

\subsection*{Special case: \boldmath $p=\infty$}

Assuming that $a\geq0$ and $b\geq0$, $\lim_{p\to\infty}(a^p+b^p)^{1/p}=\max(a,b
)$. Thus, for $p=\infty$, \eqref{e13} becomes:
\begin{eqnarray}
\int_{x=0}^{1}\!\!\!dx\,\Big\{\left[h'(x)+h'(t)\right]\max\Big(|x-t|,h(x)+h(t)
\Big)+\left[h'(x)-h'(t)\right]\max\Big(|x-t|,\left|h(x)-h(t)\right|\Big)
\nonumber\\
-\left[h'(x)+h'(t)\right]\max\Big(x+t,\left|h(x)-h(t)\right|\Big)
-\left[h'(x)-h'(t)\right]\max\Big(x+t,h(x)+h(t)\Big)\nonumber\\
+\left[1+h'(x)h'(t)\right]\max\Big(h(t)-x,h(x)+t
\Big)+\left[1-h'(x)h'(t)\right]\max\Big(h(t)+x,h(x)+t\Big)\nonumber\\
+\left[h'(x)h'(t)-1\right]\max\Big(h(t)-x,h(x)-t\Big)-\left[1+h'(x)h'(t)\right]
\max\Big(h(t)+x,h(x)-t\Big)\Big\}=0\,.
\eqlabel{e25}
\end{eqnarray}

The first step in simplifying \eqref{e25} is to remove the max conditions. Since
$x$ and $t$ are between $0$ and $1$, we know that $|x-t|\leq1$ and $x+t\leq2\leq
h(x)+h(t)$. Moreover, if $x>t$, then $|x-t|=x-t$ and $|h(x)-h(t)|=h(t)-h(x)$.
The function $x+h(x)$ is increasing because the derivative of this function with
respect to $x$ is positive. [This is because $h'(x)$ is negative but less than 1
in absolute value in the region $0\leq x<1$; see \figref{f00}.] Hence, $x+h(x)
\geq t+h(t)$, which implies that $x-t\geq h(t)-h(x)$ when $x\geq t$. A similar
argument for the case $x\leq t$ gives $|x-t|\geq|h(x)-h(t)|$ for all $x$ and $t$
in the interval $[0,1]$. Also, since $|x-t|\geq\left|h(x)-h( t)\right|$, we have
$x+t\geq|h(x)-h(t)|$. We obtain
\begin{eqnarray}
\int_{x=0}^1\!\!\!dx\,\Big\{\left[h'(x)+h'(t)\right]\left[h(x)+h(t)\right]
+\left[h'(x)-h'(t)\right]|x-t|\nonumber\\
-\left[h'(x)+h'(t)\right](x+t)-\left[h'(x)-h'(t)\right]\left[h(x)+h(t)\right]
\nonumber\\
+\left[1+h'(x)h'(t)\right]\max\Big(h(t)-x,h(x)+t\Big)+\left[1-h'(x)h'(t)\right]
\max\Big(h(t)+x,h(x)+t\Big)\nonumber\\
+\left[h'(x)h'(t)-1\right]\max\Big(h(t)-x,h(x)-t\Big)-\left[1+h'(x)h'(t)\right]
\max\Big(h(t)+x,h(x)-t\Big)\Big\}=0\,.\nonumber
\end{eqnarray}

Next we simplify the second half of the equation. We know that $x+h(x)$ and $x-h
(x)$ are increasing functions of $x$. Therefore, the inequality $x\geq t$ is
equivalent to both conditions $h(x)-t\geq h(t)-x$ and $x+h(t)\geq t+h(x)$ 
independently. Also, for all $x$ and $t$ it is true that $ h(x)+t\geq h(t)-x$
and $h(t)+x\geq h(x)-t$. We obtain
\begin{eqnarray}
\int_{x=0}^1\!\!\!dx\,\left[h'(x)+h'(t)\right]\left[h(x)+h(t)\right]
+\int_{x=0}^1\!\!\!dx\,\left[h'(x)-h'(t)\right]|x-t|\nonumber\\
-\int_{x=0}^1\!\!\!dx\,\left[h'(x)+h'(t)\right](x+t)
-\int_{x=0}^1\!\!\!dx\,\left[h'(x)-h'(t)\right]\left[h(x)+h(t)\right]\nonumber\\
+\int_{x=0}^1\!\!\!dx\,\left[1+h'(x)h'(t)\right]\left[h(x)+t\right]
+\int_{x=0}^t\!\!\!dx\,\left[1-h'(x)h'(t)\right]\left[h(x)+t\right]\nonumber\\
+\int_{x=t}^1\!\!\!dx\,\left[1-h'(x)h'(t)\right]\left[h(t)+x\right]
+\int_{x=0}^t\!\!\!dx\,\left[h'(x)h'(t)-1\right]\left[h(t)-x\right]\nonumber\\
+\int_{x=t}^1\!\!\!dx\,\left[h'(x)h'(t)-1\right]\left[h(x)-t\right]
-\int_{x=0}^1\!\!\!dx\,\left[1+h'(x)h'(t)\right]\left[h(t)+x\right] \Big\}=0\,.
\nonumber
\end{eqnarray}

Expanding and combining terms, we get
\begin{eqnarray}
4h'(t)\int_{x=0}^1\!\!\!dx\,h(x)-t^2 h'(t)-2th(t)+4\int_{x=0}^t\!\!\!dx\,h(x)
+h'(t)\left[h(t)\right]^2-2h'(t)=0\,,
\nonumber
\end{eqnarray}
and finally using $f(t)$ instead of $h(t)$, we get
\begin{eqnarray}
4f''(t)\left[f(1)-f(0)\right] -t^2 f''(t) -2tf'(t)
+4\left[f(t)-f(0)\right] +f''(t)\left[f'(t)\right]^2 -2f''(t)=0\,.
\eqlabel{e26}
\end{eqnarray}
The solution to this differential equation that satisfies the boundary
conditions in \eqreftwo{e23}{e24} is graphed in \cite{KarpMcWo75}.

\secput{conclusion}{Discussion and Conclusions}

The nonlinear integral equation in \eqref{e14} defines the exact boundary curve
of the optimal blob that satisfies the variational problem posed in this paper.
While this integral equation is compact enough to be displayed on one line, it
is only a global characterization of the solution curve $w(x)$. Although we have
tried hard to do so, we are unable to convert this integral equation to a local
(finite-order) differential-equation characterization for $w(x)$ except for the
three cases $p=1$, $p=2$, and $p=\infty$. It is interesting that these three
values of $p$ represent the three most commonly used $L_p$ metrics; namely, the
Manhattan metric, the Euclidean metric, and the maximum metric. It would be a
remarkable technical advance if the integral equation could be transformed to
simpler equation for $w(x)$ for other values of $p$.

There are many ways to generalize the continuum problem solved in this paper.
One can allow the blob to have holes or not to be simply connected. Furthermore,
one can try to solve the analog of this problem for the case in which the
dimension of space is not 2. Finally, one can try to use the exact solution of
the continuum problem as a step in solving the discrete problem posed in the
introduction to this paper. The continuum solution provides a good approximation
to the solution of the discrete problem. The question is whether we can use the
continuum solution to construct a polynomial-time algorithm for solving the
discrete problem.

\section*{Acknowledgments}

As an Ulam Scholar, CMB receives financial support from the Center for Nonlinear
Studies at the Los Alamos National Laboratory and he is supported in part by a
grant from the U.S. Department of Energy. MAB is supported by US National 
Science Foundation Grants
CCF~0621439/0621425,  
CCF~0540897/05414009, 
and CNS 0627645.      

\bibliographystyle{plain}

\end{document}